\documentclass[journal]{IEEEtran}

\usepackage{graphics} 
\usepackage{epsfig} 
\usepackage{mathptmx} 
\usepackage{times} 
\usepackage{amsmath} 
\usepackage{amssymb}  
\usepackage{cite}
\usepackage{subfig}
\usepackage{multirow}

\newtheorem{corollary}{Corollary}
\newtheorem{theorem}{Theorem}
\newtheorem{remark}{Remark}

\title{Computing Critical $k$-tuples in Power Networks}

\author{Kin Cheong Sou, Henrik Sandberg and Karl Henrik Johansson\thanks{The authors are with the ACCESS Linnaeus Center and the Automatic Control Lab, the School of Electrical Engineering, KTH Royal Institute of Technology, Sweden. {\tt \small \{sou,hsan,kallej\}@kth.se} \newline This work is supported by the European Commission through the VIKING project, the Swedish Foundation for Strategic Research (SSF) and the Knut and Alice Wallenberg Foundation.}}

\begin{document}

\maketitle

\begin{abstract}
  In this paper the problem of finding the sparsest (i.e., minimum cardinality) critical $k$-tuple including one arbitrarily specified measurement is considered. The solution to this problem can be used to identify weak points in the measurement set, or aid the placement of new meters. The critical $k$-tuple problem is a combinatorial generalization of the critical measurement calculation problem. Using topological network observability results, this paper proposes an efficient and accurate approximate solution procedure for the considered problem based on solving a minimum-cut (Min-Cut) problem and enumerating all its optimal solutions. It is also shown that the sparsest critical $k$-tuple problem can be formulated as a mixed integer linear programming (MILP) problem. This MILP problem can be solved exactly using available solvers such as CPLEX and Gurobi. A detailed numerical study is presented to evaluate the efficiency and the accuracy of the proposed Min-Cut and MILP calculations.
\end{abstract}

\section*{Notation}
\begin{tabular}{ll}
  $\mathcal{A}$ & A subset of transmission lines whose removal \\
  & partitions the network into two disjoint parts. \\
  $\mathcal{E}$ & The set of all transmission lines. \\
  $\mathcal{G}$ & The graph of the power network. \\
  $H$ & The Jacobian of the measurement function. \\
  $H(I,J)$ & A submatrix of $H$, consisting of the rows and\\
  & columns in the index sets $I$ and $J$ respectively. \\
  $H(i,:)$ & The $i^{\rm th}$ row of matrix $H$. \\
  $\bar{I}$ & The complement of an index set $I$. \\
  $\bar{j}$ & The complement of a singleton set $\{j\}$. \\
  $M$ & A large scalar constant treated as ``infinity'' \\
  & in the MILP procedure. \\
  $m$ & The number of measurements (rows of $H$). \\
  $n$ & The number of buses (columns of $H$). \\
  $S$ & A subset of buses (i.e., $S \subset \mathcal{V}$). \\
  & $S$ is used to define a partition of the network. \\
  $v_i$ & Node weights for bus $i$. \\
  $w_{ij}$ & Edge weights for transmission line $\{i,j\}$. \\
  $\tilde{w}_{ij}$ & Edge weights which also account for the bus\\
  & weights connected to $\{i,j\}$. $\tilde{w}_{ij} = w_{ij}+v_i+v_j$. \\
  $\mathcal{V}$ & The set of all buses. \\
  $\theta$ & The $n \times 1$ state vector (phase angles). \\
  $\delta(S)$ & Cut capacity of $S$. The sum of all weights of \\
  & the edges cut by the partition defined by $S$. \\
  $\delta_{\rm inj}(S)$ & The sum of all weights of the buses connected \\
  & by edges which are cut by the partition of $S$. \\
  $\tilde{\delta}(S)$ & Modified cut capacity, where the edge weights \\
  & $w_{ij}$ are replaced by $\tilde{w}_{ij}$.
\end{tabular}

\section{Introduction}

\subsection{Critical $k$-tuples}

A modern SCADA/EMS system relies heavily on the state estimator, which estimates the power network states (e.g., the phase angles of bus voltages) based on measurements such as transmission line power flows, bus power injections and bus voltages. An important question related to state estimation is whether the network is observable or not; whether the states can be uniquely determined based on the available measurements. This is a central issue of network observability analysis (e.g., \cite{Monticelli_Wu85,KCD80,Abur_Exposito_SEbook,Monticelli_SEbook,AAG08,Gou06,Pruneda2010277,solares:336,CCPS06} and the references therein). While the measurements are typically placed so that a power network is observable, there exist weak points known as critical measurements. By definition, if a critical measurement is lost (e.g., failure of a meter), then the network becomes unobservable (i.e., the states can no longer be uniquely determined). The notion of critical measurement also plays an important role in another vital power network state estimation function, namely, bad data detection (e.g., \cite{Monticelli_Wu_Yen86,Clements_Davis86,MVCRP85,Abur_Exposito_SEbook,Monticelli_SEbook} and the references therein). Specifically, a bad data detection scheme based on measurement residual cannot identify whether a meter is faulty or not if the corresponding measurement is critical. A generalization of the concept of critical measurement is a critical $k$-tuple, where $k$ is any natural number. A critical $k$-tuple is a set of $k$ measurements such that if all measurements in the set are lost then the network becomes unobservable. However, losing any subset of $p < k$ measurements would not result in the loss of observability. A critical $2$-tuple is also referred to as a critical set, where bad data can be detected but not identified (other terminologies include minimally dependent set or bad data group, e.g., \cite{KC91,AAG09,AH86}). Critical $k$-tuples of larger cardinalities are also of practical interest, as will be explained later. How to compute them is the main topic of this paper.


While critical $k$-tuple, network observability, and bad data detection are closely related, the mathematical tools to analyze them are different. While critical $k$-tuples provide the measurements to remove to render the network unobservable, the topic of network observability is the opposite. They include checking whether a network is observable or not, and in case of an unobservable network which parts of it is still observable (i.e., finding observable islands). Likewise, critical $k$-tuple computation and bad data detection are separate issues: Essentially, the measurement residual based bad data detection theory investigates what detection can be achieved beyond the limitations imposed by critical measurements or critical $k$-tuples. These techniques are, consequently, not concerned with finding the critical $k$-tuples.


Techniques to identify critical $1$-tuples (i.e., critical measurements) and critical $2$-tuples (i.e., critical sets) are widely known (e.g., \cite{Clements_Davis86,Abur_Exposito_SEbook,CKD81,KC91,AAG09,AH86}). In \cite{LondonJr2004583} and \cite{LAB00} the calculation of critical $k$-tuples for $k > 2$ is considered. However, the procedure in \cite{LAB00} is efficient only for finding critical $k$-tuples of lower cardinalities (i.e., $k \le 3$), as will be explained and numerically demonstrated later in this paper. The computation of critical $k$-tuples is inherently computationally intensive, because finding critical $k$-tuples amounts to a combinatorial search, as will be discussed in detail in this paper.

\subsection{Problem Formulation and its Motivation}


The general setup of this paper is the standard state estimation problem over a linearized DC power flow network \cite{Abur_Exposito_SEbook,Monticelli_SEbook}. The particular problem considered is to find the sparsest (i.e., minimum cardinality) critical $k$-tuples involving at least one arbitrarily specified measurement. By parameterizing the sparsest critical $k$-tuple problem with the specified measurement, it is possible to examine all weak points in the network and not just the weakest point at the boundary of the network. The precise description of the considered problem is as follows. Let $m$ be the number of measurements in the power network, and $n$ be the number of states (i.e., the phase angles of the bus voltage phasors). It is assumed that $m > n$. Let $H \in \mathbb{R}^{m \times n}$ be the Jacobian of the state-to-measurement function in a linearized model. Denote $H(I,J)$ as the submatrix of $H$ formed by including the rows in an index set $I$ and the columns in an index set $J$. Also denote $\bar{I}$ and $\bar{J}$ as the complements of $I$ and $J$, respectively. Then according to, for instance \cite{Monticelli_SEbook} (Theorem p.165), the measurements in an index set $I$ form a critical $k$-tuple if and only if ${\rm rank}(H(\bar{I},\bar{j})) < n-1$ for any $j$. Here $j$ is the index of an arbitrary reference bus, and $\bar{j}$ denotes the set of all indices except $j$. The sparsest critical $k$-tuple problem for a specified measurement $i$ can be written as:
\begin{equation} \label{opt:sckt}
  \begin{array}{rl}
    \mathop{\rm minimize}\limits_{I} & {\rm card}(I) \\
    \textrm{subject to} & {\rm rank}(H(\bar{I},\bar{j})) < n - 1 \vspace{1mm} \\
    & i \in I
  \end{array}
\end{equation}
where ${\rm card}(\cdot)$ denotes either the cardinality of a set or the number of nonzero entries of a vector, depending on the input argument. Notice that (\ref{opt:sckt}) does not explicitly impose the condition that $I$ cannot contain any strictly proper subsets whose removal makes $H$ rank deficient. However, this condition is always satisfied at optimality. Problem (\ref{opt:sckt}) requires a combinatorial search of the rows whose removal makes the $H$ matrix rank deficient. In general, no efficient algorithm is available to exactly solve (\ref{opt:sckt}). However, specializing (\ref{opt:sckt}) to the case of power system state estimation results in a significant solution efficiency gain because of the special structure of $H$. The demonstration of this is the main contributions of the paper. Problem (\ref{opt:sckt}) is motivated from the following applications:

\subsubsection{Identifying measurements in small cardinality critical $k$-tuples}
While not directly solved as an optimization problem, (\ref{opt:sckt}) is addressed in \cite{LAB00} (Definition 1, even though the term ``critical set'' in \cite{LAB00} has a different meaning than the one here). For any given measurement set, \cite{LAB00} finds the measurements such that optimal objective value of (\ref{opt:sckt}) is less than or equal to three. This information is used to determine the set of additional measurements to be included, so that the network becomes more robust to meter failures. Other, but related, meter inclusion problems are also considered in \cite{CCPSM08,GA01}. This paper, on the other hand, solves (\ref{opt:sckt}) for all $i$, regardless of the corresponding optimal objective value. These include the measurements in the critical $k$-tuples of cardinalities less than or equal to three as in \cite{LAB00}. However, the information of the sparsest critical $k$-tuples of larger cardinalities can be used for a measurement inclusion scheme with a more stringent robustness requirement.

\subsubsection{Planning of measurement sets}
Instead of expanding a pre-existing measurement set as in \cite{LAB00,CCPSM08,GA01}, it is possible to obtain a cost effective yet meter failure robust measurement set by removing appropriate measurements from the full set. The solution to (\ref{opt:sckt}) with $H$ corresponding to the full measurement set can provide insight into which measurements can be removed. This measurement removal strategy has the advantage that it does not assume any pre-existing measurement set which can affect the final measurement set. In Section~\ref{subsec:meter_placement} a numerical example is presented to demonstrate the potential of the above measurement removal scheme.

\subsubsection{Cyber-security of power networks}
Based on a result in \cite{KJTT10} (Corollary~\ref{rem:equivalence} in Section~\ref{subsec:equivalence}), problem (\ref{opt:sckt}) is equivalent to another cardinality minimization problem: (\ref{opt:suoa}) to be described in Section~\ref{subsec:equivalence}. Problem~(\ref{opt:suoa}) arises from cyber-security analysis of power networks (e.g., \cite{LRN09,BRWKNO10,IR-EE-RT_2010:071,DS_SGC2010,KJTT10}). In particular, \cite{IR-EE-RT_2010:071,DS_SGC2010,SSJ_CDC2011_mincut} analyze the vulnerability of each measurement $i$ using (\ref{opt:suoa}), where a malicious attacker inflicts ``bad data'' in a critical $k$-tuple. In this case, sparsest critical $k$-tuples of larger cardinalities (i.e., $> 3$) are of interest because the ``bad data'' is intentional instead of occurring by chance. The solution of (\ref{opt:suoa}) can be used to identify the weak points in the measurement set in the cyber-security setting.

Finally, note that for complete safeguard against bad data or cyber-attack, the set of all critical $k$-tuples (in addition to the sparsest ones found by solving (\ref{opt:sckt})) should be computed. However, this would require an enumeration which is not computationally tractable for realistic applications. The calculation of the sparsest critical $k$-tuples in (\ref{opt:sckt}) can identify the network vulnerabilities, subject to practical computation constraints.


\subsection{Contributions and Related Work} \label{subsec:contributions}

This paper presents two methods to solve (\ref{opt:sckt}). The first method is efficient but suboptimal. It utilizes a sufficient condition for critical $k$-tuples candidates in \cite{KJTT10, KCD80}. This condition is topological. In the setting of this paper, the condition states that for any set of transmission lines whose cut would separate the network into two disjoint parts, removing all line and injection measurements associated with these transmission lines would make the network unobservable. Using this sufficient condition, a restricted version of the sparsest critical $k$-tuple problem in (\ref{opt:sckt}) can be stated as follows. If the specified measurement $i$ is a line power flow, then the corresponding transmission line must be cut. On the other hand, if the specified measurement is a power injection at a bus, then one of the incident transmission lines must be cut. Then the rest of the transmission lines are cut (or not cut) in order to minimize the number of measurements removed, while dividing the network into two parts. This cut problem, while being a restricted version of (\ref{opt:sckt}), is still combinatorial. However, if the injection measurements are \emph{not directly} counted towards the optimization objective (to be made precise later), then this modified problem becomes a classical minimum cut problem (Min-Cut) (e.g., \cite{BT97}). Min-Cut admits scalable solution algorithms (e.g., \cite{Stoer:1997:SMA:263867.263872,FF_MAX-FLOW}). The solution to the Min-Cut problem can be used as a suboptimal solution to (\ref{opt:sckt}). In fact, due to \cite{springerlink:10.1007/BFb0120902,Schrage_Baker_1978,Lawer_1979}, it is possible to efficiently enumerate all optimal solutions to the Min-Cut problem and pick the best available suboptimal solutions to (\ref{opt:sckt}). This is the idea of the first method of this paper.

Two previous results are related to the first proposed method. As mentioned before, \cite{LAB00} addresses (\ref{opt:sckt}). In \cite{LAB00}, a (non-unique) set of measurements is chosen to be the \emph{basic measurements}. Then the critical $k$-tuples containing exactly one basic measurement can be identified using a matrix factorization approach generalizing the one in \cite{Abur_Exposito_SEbook} (Chapter 4.5.4). To find critical $k$-tuples containing more than one basic measurements, a recursive application of the matrix factorization approach for finding critical $k$-tuples with only one basic measurement is required. For larger $k$, the recursion becomes more expensive as there are ${n \choose p}$ possible combinations of $p$ basic measurements to be included in the critical $k$-tuples, for different $p \le k$. To solve (\ref{opt:sckt}) for all possible $i$, in total $\sum\limits_{p=1}^n {n \choose p}$ applications of the matrix factorization procedure are required. The computation effort is exponential in terms of network size (i.e., the number of buses $n$). In summary, \cite{LAB00} is accurate but the procedure is efficient only for a sparse measurement set (so that critical $k$-tuples of high cardinalities will not be encountered). The proposed method in this paper, on the other hand, is efficient for solving (\ref{opt:sckt}) irrespective of the cardinality of the critical $k$-tuple, because (\ref{opt:sckt}) is approximately solved via Min-Cut. However, as it will be numerically demonstrated in Section~\ref{sec:numerical_experiment}, the accuracy of the proposed method suffers when the measurement set becomes sparse. In this sense, \cite{LAB00} and the proposed method are complementary to each other. Another closely related work is \cite{KJTT10}, which considers a variant of (\ref{opt:sckt}). In this variant, the sparsest critical $k$-tuple also contains at least one measurement. However, instead of being user-specified, this measurement is chosen by the optimization to find the sparsest non-empty critical $k$-tuple. Solving (\ref{opt:sckt}) for all $i$ leads to the solution to the problem in \cite{KJTT10} but the converse is not true. In addition, \cite{KJTT10} does not pose their problem as (\ref{opt:sckt}) defined in this paper. Most importantly, the problem in \cite{KJTT10} is posed as a submodular function minimization problem \cite{Cunningham85submodularfunction,Mccormick07submodularfunction}. While theoretically polynomial-time algorithms exist for solving this problem (notably the ellipsoid method \cite{BGT81} and more recently \cite{Iwata:2009:SCA:1496770.1496903} of which the complexity is $O(m^8 \log(m))$), no practically efficient algorithms for this class of problem have been observed. On the other hand, the Min-Cut problem encountered by the proposed method can be solved efficiently both in theory and in practice. For example, the complexity of \cite{Stoer:1997:SMA:263867.263872} is $O(mn+n^2\log(n))$. The practical efficiency will be demonstrated by the numerical experiment later in this paper.


The second proposed method, based on mixed integer linear programming (MILP), is exact under a mild assumption, but it is less time-efficient. The method is based on the equivalence between (\ref{opt:sckt}) and (\ref{opt:suoa}), to be described in Section~\ref{subsec:equivalence}. This means that (\ref{opt:sckt}) can be solved by instead solving (\ref{opt:suoa}). Previous attempts to approximately solve (\ref{opt:suoa}) include, for instance, \cite{LRN09} describing an attempt to use matching pursuit (e.g., \cite{Mallat93matchingpursuit}), and \cite{STJ_SCS2010} about the application of LASSO \cite{LASSO}. The MILP formulation, based on \cite{SSJ_CDC2011_mincut}, does not admit any polynomial time solution algorithms in general. However, there exist good MILP solvers such as CPLEX \cite{CPLEX} or Gurobi \cite{Gurobi}. The major novelty of this second contribution of the paper is the combination of \cite{KJTT10} and \cite{SSJ_CDC2011_mincut}.

\subsection{Organization of the Paper}


The rest of the paper is organized as follows. In Section~\ref{sec:background} three known theorems from \cite{KCD80} and \cite{KJTT10} are reviewed, and a corollary is derived. These theorems form the theoretical foundation of this paper. Section~\ref{sec:background} also reviews the Min-Cut problem, which is an important part of the proposed algorithm. Section~\ref{sec:Min_Cut} describes the first contribution of the paper: a Min-Cut based algorithm which makes use of the topological characterization of network observability to find the sparsest critical $k$-tuples. In Section~\ref{sec:MILP} the second contribution, the exact MILP formulation, is derived with some properties discussed. In Section~\ref{sec:numerical_experiment} some case studies are presented to evaluate the performance of the proposed algorithms. Finally Section~\ref{sec:conclusion} concludes the paper.


\section{Technical Background} \label{sec:background}

This section reviews some known results needed for the derivation of the contributions of this paper. Theorems adopted from known sources are stated without proof.

\subsection{A Topological Sufficient Condition for Critical $k$-tuple Candidates}
The first statement is adopted from \cite{KCD80} (Theorem 5). It provides a sufficient and necessary condition for network observability in terms of spanning trees, which are loop-free connected subgraphs of the power network retaining all buses but subsets of the transmission lines.
\begin{theorem} \label{thm:spanning_tree}
  A power network is observable if and only if there exists a spanning tree with an assignment function, mapping from the set of the transmission lines in the spanning tree to the set of line power flow and injection measurements of the original power network. The assignment function satisfies the following properties:
  \begin{enumerate}
    \item Two distinct spanning tree transmission lines map to two distinct measurements.
    \item If the line power flow of a spanning tree transmission line is measured, then this transmission line maps to its own line measurement under the assignment function.
    \item If the line power flow of a spanning tree transmission line is not measured, then the injection measurement of one of the two terminal buses of this transmission line is the value of the assignment function.
  \end{enumerate}
\end{theorem}

The following theorem, which is the main theoretical basis of this paper, is adopted from Theorem 2 of \cite{KJTT10}. It states that if an appropriate choice of measurements are removed, then it becomes impossible to form any spanning tree with an assignment function defined as in Theorem~\ref{thm:spanning_tree}. Hence, the statement provides a sufficient condition for finding candidates for critical $k$-tuples.
\begin{theorem} \label{thm:ckt}
  Let $\mathcal{A}$ be any set of transmission lines whose cut would divide the power network into two disjoint parts. Then removing all line power flow measurements in $\mathcal{A}$ and all power injection measurements of the buses connected by the lines in $\mathcal{A}$ would render the power network unobservable.
\end{theorem}
\begin{remark}
  Any spanning tree of the network contains at least one transmission line in $\mathcal{A}$. However, under the measurement removal scheme in Theorem~\ref{thm:ckt}, it is impossible to define any assignment function in Theorem~\ref{thm:spanning_tree} for this line. Hence the network becomes unobservable.
\end{remark}
\begin{remark} \label{rem:ckt}
  While Theorem~\ref{thm:ckt} provides the sets of measurements whose removal would render the network unobservable, these sets are not necessarily critical $k$-tuples since their subsets might also render the network unobservable.
\end{remark}
\begin{remark}
  The original version of Theorem~\ref{thm:ckt}, as in \cite{KJTT10}, is more general in that it allows the situations in which $\mathcal{A}$ divides the network into more than two disjoint parts. However, the method proposed in this paper cannot exploit the additional generality.
\end{remark}

\subsection{Sparsest Critical $k$-tuple Problem as a Cardinality Minimization Problem} \label{subsec:equivalence}
The following cardinality minimization problem has been studied in power network cyber-security (e.g., \cite{LRN09,DS_SGC2010}):
\begin{equation} \label{opt:suoa}
  \begin{array}{rl}
    \mathop{\rm minimize}\limits_{\theta} & {\rm card}(H \theta) \\
    \textrm{subject to} & H(i,:) \theta = 1 \vspace{1mm}
  \end{array}
\end{equation}
The following theorem, adopted from Theorem 1 in \cite{KJTT10}, establishes that the sparsest critical $k$-tuple problem in (\ref{opt:sckt}) is equivalent to (\ref{opt:suoa}).
\begin{theorem} \label{thm:sckt_sa}
   An index set $I$ is a feasible solution to (\ref{opt:sckt}) if and only if there exists a feasible solution $\theta$ in (\ref{opt:suoa}) such that $H(j,:) \theta = 0$ whenever $j \notin I$.
\end{theorem}
Theorem~\ref{thm:sckt_sa} implies the following statement (proved in Appendix) establishing the equivalence between (\ref{opt:sckt}) and (\ref{opt:suoa}).
\begin{corollary} \label{rem:equivalence}
  The optimization problems in (\ref{opt:sckt}) and (\ref{opt:suoa}) are equivalent in that $\theta^\star$ is an optimal solution to (\ref{opt:suoa}) if and only if $I^\star = \{j \;\; \vline \;\; H(j,:) \theta^\star \neq 0 \}$ is an optimal solution to (\ref{opt:sckt}).
\end{corollary}

\subsection{Min-Cut Problem on an Undirected Graph}
Consider an undirected graph $\mathcal{G} = (\mathcal{V},\mathcal{E})$ where $\mathcal{V}$ and $\mathcal{E}$ denote the set of nodes and the set of edges respectively, and let each edge $\{i,j\} \in \mathcal{E}$ be weighted with a scalar $w_{ij}$. Let $S \subset \mathcal{V}$ be any subset of $\mathcal{V}$. Define the cut capacity function
\begin{equation} \label{eqn:cut_capacity}
\begin{array}{ll}
   \delta(S) \triangleq \sum\limits_{\{i,j\} \in \mathcal{E}} w_{ij} & \textrm{such that either (a) $i \in S$ and $j \notin S$} \\ & \textrm{or (b) $i \notin S$ and $j \in S$.}
\end{array}
\end{equation}
For any two distinct nodes $s$ and $t$, the $s-t$ Min-Cut problem seeks to find a partition of $\mathcal{V}$ into $\mathcal{V} = S \cup (\mathcal{V} \setminus S)$ such that $s$ and $t$ are in different partitions, and the cut capacity is minimized:
\begin{equation} \label{opt:min-cut}
  \begin{array}{cl}
    \mathop{\rm minimize}\limits_{S \subset \mathcal{V}} & \delta(S) \\
    \textrm{subject to} & s \in S \quad {\rm and} \quad t \notin S
  \end{array}
\end{equation}
For more detail regarding the Min-Cut problem in (\ref{opt:min-cut}), see for example \cite{BT97}. For efficient solution algorithms, see for example \cite{Stoer:1997:SMA:263867.263872,FF_MAX-FLOW}. The Min-Cut problem is a subproblem to be solved in the proposed critical $k$-tuple calculation algorithm to be described in the next section.

\section{Approximate Critical k-tuple calculation via Min-Cut Optimization} \label{sec:Min_Cut}

\subsection{A Graph-Oriented Optimization Problem Related to (\ref{opt:sckt})} \label{subsec:GOOP}
The sufficient condition in Theorem \ref{thm:ckt} provides a topological characterization of a subset of the solution candidates of the sparsest critical $k$-tuple problem in (\ref{opt:sckt}). This characterization leads to a graph-oriented optimization problem which is related to, but not exactly the same as, (\ref{opt:sckt}). The development is as follows. Denote the power network as $\mathcal{G} = (\mathcal{V},\mathcal{E})$, where $\mathcal{V}$ is the set of all buses and $\mathcal{E}$ is the set of all transmission lines. Then the set $\mathcal{A}$ in Theorem~\ref{thm:ckt}, whose cut would partition $\mathcal{G}$, can be characterized by a bus subset $S \subset \mathcal{V}$ such that
\begin{equation} \label{eqn:SA}
\mathcal{A} = \big\{\{i,j\} \; \vline \;\; \textrm{either (a) $i \in S$ and $j \notin S$ or (b) $i \notin S$ and $j \in S$}\big\}
\end{equation}
To describe the number of removed measurements associated with $S$ (i.e., $\mathcal{A}$) according to Theorem~\ref{thm:ckt}, the following definitions are required. Let $w_{ij}$ be the number of meters on a transmission line $\{i,j\} \in \mathcal{E}$, and $v_j$ be the number of injection flow meters on a bus $j \in \mathcal{V}$. Then associated with $S$, the number of line power flow measurements to remove is $\delta(S)$ defined in (\ref{eqn:cut_capacity}). In addition, the number of power injection measurements to remove can be defined as
\begin{displaymath}
\begin{array}{l}
  \delta_{\rm inj}(S) \triangleq \sum\limits_{j \; \in \mathcal{V}} v_j \quad \textrm{such that either} \vspace{1mm} \\
  \begin{array}{r} \textrm{(a) $j \in S$ and $\exists \; i \in (\mathcal{V} \setminus S)$ s.t.\ $\{i,j\} \in \mathcal{E}$} \vspace{1mm} \\
  {\rm or} \quad \textrm{(b) $j \in (\mathcal{V} \setminus S)$ and $\exists \; i \in S$ s.t.\ $\{i,j\} \in \mathcal{E}$} \vspace{1mm}
  \end{array}
\end{array}
\end{displaymath}
Hence, associated with $S$, the total number of measurements to be removed is $\delta(S) + \delta_{\rm inj}(S)$. Lastly, the constraint in (\ref{opt:sckt}) that one specified measurement must be included in the critical $k$-tuple should be enforced. For simplicity of discussion, for the moment it is assumed that the specified measurement is a line power flow. The case of power injection will be handled in the end of this section. Now suppose the specified line power flow meter $i$ is on transmission line $\{s,t\}$, then the corresponding topological constraint is that $s \in S$ and $t \notin S$. In summary, the graph oriented optimization problem, set up as an approximation to (\ref{opt:sckt}), is described as:
\begin{equation} \label{opt:sckt2}
  \begin{array}{cl}
    \mathop{\rm minimize}\limits_{S \subset \mathcal{V}} & \delta(S) + \delta_{\rm inj}(S) \\
    \textrm{subject to} & s \in S \quad {\rm and} \quad t \notin S
  \end{array}
\end{equation}
Strictly speaking an optimal solution to (\ref{opt:sckt2}) is a set of buses, with the corresponding set of ``cut'' transmission lines defined in (\ref{eqn:SA}). However, it is more convenient to treat the optimal solution as the corresponding set of measurements to be removed, as prescribed by Theorem~\ref{thm:ckt}. Solving (\ref{opt:sckt2}) yields a sparse set measurements whose removal makes the network unobservable. The numerical experiment in Section~\ref{sec:numerical_experiment} will demonstrate the usefulness of (\ref{opt:sckt2}). However, it should be emphasized that an optimal solution to (\ref{opt:sckt2}) is not necessarily a sparsest critical $k$-tuple in (\ref{opt:sckt}). The reason is twofold. First, since Theorem~\ref{thm:ckt} is a sufficient condition, (\ref{opt:sckt2}) searches only through a subset of the sets of measurements whose removal would render the network unobservable. Second, as pointed out in Remark~\ref{rem:ckt}, an optimal solution to (\ref{opt:sckt2}) does not even need to be a critical $k$-tuple. These restrictions will be demonstrated by the numerical experiment in Section~\ref{sec:numerical_experiment}.

\subsection{Min-Cut Approximate Solution Procedure for (\ref{opt:sckt})}
\begin{figure}
  \begin{center}
    \includegraphics[scale=0.45]{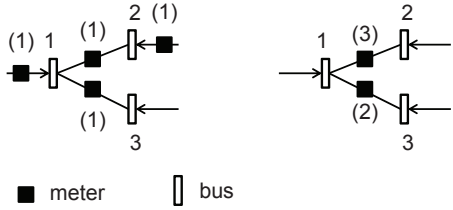}
  \end{center}
  \caption{Illustration of the modified cut capacity function. Left: Metering scenario in (\ref{opt:sckt2}). Right: Transmission line metering scenario pertaining to $\tilde{\delta}(S)$ in (\ref{opt:min-cut_mod}). The number in the parenthesis indicates the number of times a meter is repeated. If both 1-2 and 1-3 are cut, the true cost $\delta(S)+\delta_{\rm inj}(S)$ is 4. However, the approximate cost $\tilde{\delta}(S)$ is 5.}
  \label{fig:inj2tl}
\end{figure}

Compared with the tractable Min-Cut problem in (\ref{opt:min-cut}), (\ref{opt:sckt2}) is a combinatorial optimization problem because of the additional term $\delta_{\rm inj}(S)$ in the objective function. To overcome the computational difficulty, it is proposed in this paper that $\delta_{\rm inj}(S)$ is indirectly accounted for by solving the following Min-Cut problem:
\begin{equation} \label{opt:min-cut_mod}
  \begin{array}{cl}
    \mathop{\rm minimize}\limits_{S \subset \mathcal{V}} & \tilde{\delta}(S)\\
    \textrm{subject to} & s \in S \quad {\rm and} \quad t \notin S,
  \end{array}
\end{equation}
where $\tilde{\delta}(S)$ is defined according to (\ref{eqn:cut_capacity}) with modified edge weights $\tilde{w}_{ij} \triangleq w_{ij} + v_{i} + v_{j}$ for all $\{i,j\} \in \mathcal{E}$, with $w_{ij}$, $v_{i}$ and $v_{j}$ defined in Section~\ref{subsec:GOOP}. $\tilde{\delta}(S)$ corresponds to a modified metering scenario where an injection meter of a bus is moved to all incident transmission line(s). However, this modification can lead to overcounting of injection meters as opposed to solving (\ref{opt:sckt2}). See Fig.~\ref{fig:inj2tl} for an illustration.
Since (\ref{opt:sckt2}) and (\ref{opt:min-cut_mod}) have the same constraint, the optimal solution to (\ref{opt:min-cut_mod}) is a suboptimal solution to (\ref{opt:sckt2}). In addition, by the results in \cite{springerlink:10.1007/BFb0120902,Schrage_Baker_1978,Lawer_1979}, it is possible to efficiently enumerate all optimal solutions to the Min-Cut problem in (\ref{opt:min-cut_mod}). Hence, the best available suboptimal solution to (\ref{opt:sckt2}) can be chosen. However, it is emphasized that the strategy of solving (\ref{opt:min-cut_mod}) can only provide a sparse set of measurements including $\{s,t\}$, and the removal of these measurements makes the network unobservable. From (\ref{opt:sckt}) to (\ref{opt:sckt2}) and then to (\ref{opt:min-cut_mod}) these two transitions induce their respective limitations. As it was explained earlier, (\ref{opt:sckt}) and (\ref{opt:sckt2}) are two different optimization problems. Moreover, solving (\ref{opt:min-cut_mod}) is not equivalent to solving (\ref{opt:sckt2}), as illustrated in Fig.~\ref{fig:inj2tl}. The quality of approximately solving (\ref{opt:sckt2}) via (\ref{opt:min-cut_mod}) depends on the ratio between the number of transmission line measurements and bus injection measurements. In the extreme case where there is no injection measurement, (\ref{opt:min-cut_mod}) is the same as (\ref{opt:sckt2}). The accuracy and efficiency of the proposed Min-Cut procedure will be numerically assessed in Section~\ref{sec:numerical_experiment}. The following algorithm summarizes the Min-Cut based approximate solution procedure for (\ref{opt:sckt}), where the specified measurement is a line power flow on a transmission line.

\noindent {\bf Algorithm 1: Min-Cut procedure for transmission line case:}
\begin{description}
\item[Step 1] \hfill \\
In the power network graph $\mathcal{G} = (\mathcal{V},\mathcal{E})$, define arc weights $\tilde{w}_{ij}$ as the number of meters on a transmission line $\{i,j\} \in \mathcal{E}$, plus the number of meters on buses $i$ and $j$.
\item[Step 2] \hfill \\
Suppose the specified transmission line is $\{s,t\} \in \mathcal{E}$. Setup a $s-t$ Min-Cut problem as in (\ref{opt:min-cut_mod}). Solve (\ref{opt:min-cut_mod}) using algorithms such as \cite{Stoer:1997:SMA:263867.263872,FF_MAX-FLOW} for an optimal solution, which is a set of ``cut'' transmission lines. The line power flows measurements and injections measurements at the terminal buses constitute a suboptimal solution to the sparest critical $k$-tuple problem in (\ref{opt:sckt}).
\item[Step 3] \hfill \\
 Use the results in \cite{springerlink:10.1007/BFb0120902,Schrage_Baker_1978,Lawer_1979} to enumerate all optimal solutions to (\ref{opt:min-cut_mod}). Pick the best suboptimal solution to (\ref{opt:sckt}) among all optimal solutions to (\ref{opt:min-cut_mod}).
\end{description}
Even the best available suboptimal solution to (\ref{opt:sckt}) might not be a critical $k$-tuple in that there might be a strictly proper subset whose removal makes the network unobservable. To make sure a critical $k$-tuple is obtained, an enumeration is required to see which measurements in the suboptimal solution can be eliminated. However, since the suboptimal solution typically contains very few measurements, the enumeration is not expensive.

In the case where the specified measurement $i$ in (\ref{opt:sckt}) is a power injection on a bus, the following procedure can be applied:

\noindent {\bf Algorithm 2: Min-Cut procedure for bus injection case:}
\begin{description}
\item[Step 1] \hfill \\
Let $\mathcal{G} = (\mathcal{V},\mathcal{E})$ be the power network graph and let $i \in \mathcal{V}$ be the bus with the considered injection measurement. For each $j \in \mathcal{V}$ such that $\{i,j\} \in \mathcal{E}$, apply Algorithm 1 on transmission line $\{i,j\}$.
\item[Step 2] \hfill \\
Among all solutions provided by Algorithm 1 applied to $\{i,j\}$, pick the one with the minimum cost in (\ref{opt:sckt}) as the best available solution to (\ref{opt:sckt}).
\end{description}
The numerical examples in Section~\ref{sec:numerical_experiment} illustrate the performance of these algorithms.

\section{Exact Sparsest Critical k-tuple Problem Formulation as a MILP Problem} \label{sec:MILP}
%
%

Theorem~\ref{thm:sckt_sa} and Corollary~\ref{rem:equivalence} state that the sparsest critical $k$-tuple problem in (\ref{opt:sckt}) can be solved by solving the cardinality minimization problem in (\ref{opt:suoa}). Problem (\ref{opt:suoa}) can be formulated as a MILP problem, as mentioned in \cite{SSJ_CDC2011_mincut}. The key to the formulation is the counting of the cardinality of vector $H \theta$. To achieve this, an additional binary decision vector $y \in \{0,1\}^m$ and a scalar constant $M > 0$ are needed. If $M$ is large enough, then the constraint
\begin{displaymath}
  |H(j,:) \theta| \le M y(j)
\end{displaymath}
provides a cardinality counting mechanism via $y$. If $|H(j,:) \theta| > 0$, then $y(j) = 1$. If $|H(j,:) \theta| = 0$, then $y(j)$ can be either 0 or 1. However, since (\ref{opt:suoa}) seeks to minimize the cardinality of $H \theta$, as it will clear shortly, $y(j)$ must be zero at optimality. The constant $M$ must be chosen large enough so that it is larger than $\max\limits_j \{|H(j,:)\theta^\star|\}$ for at least one optimal solution $\theta^\star$ of (\ref{opt:suoa}). In the special case where all line power flows are measured, the method in \cite{SSJ_CDC2011_mincut} can be used to compute $M$. In other cases, the general guideline is that $M$ should be as large as possible, before the optimization solver complains about numerical difficulties. Suppose $\theta^0$ is a typical state vector under normal operation, then $\alpha \frac{\max |H \theta^0|}{\min |H \theta^0|}$ with some $\alpha > 1$ can be a reasonable guess for $M$. The choice of $M$ is the only heuristic part of the otherwise exact sparsest critical $k$-tuple problem formulation. The MILP formulation of (\ref{opt:suoa}) is as follows:
\begin{equation} \label{opt:suoa_MILP}
  \begin{array}{cccl}
    \mathop{\rm minimize}\limits_{\theta, \; y} & \quad \sum\limits_j y(j) & & \\
    \textrm{subject to} & H \theta & \le & M y \\
    & -H \theta & \le & M y \\
    & H(i,:) \theta & = & 1 \\
    & y(j) & \in & \{0,1\} \quad \forall \; j
  \end{array}
\end{equation}
Note that since the objective function is $\sum\limits_j y(j)$, at optimality for any $j$ such that $|H(j,:) \theta| = 0$, the corresponding $y(j)$ must be zero. Hence, $\sum\limits_j y(j) = {\rm card}(H \theta)$. Finally, notice that if the measurements in a certain set $\mathcal{P}$ are considered very reliable and are immune from faults, then (\ref{opt:suoa_MILP}) can be modified accordingly by adding the constraint $y(j) = 0$ for all $j \in \mathcal{P}$.


\section{Case Study} \label{sec:numerical_experiment}
Numerical experiment results are demonstrated in this section. All computations are performed on a laptop with an Intel Core i5 2.53GHz CPU and 4GB of memory. All Min-Cut problems are solved in MATLAB using \cite{Gleich06contentsmatlab}, which calls the libraries from \cite{2002:BGL:504206}. All MILP problems are solved in MATLAB using Gurobi \cite{Gurobi} via Gurobi Mex \cite{Gurobi_Mex}.

\subsection{Comparison with the Procedure in \cite{LAB00}} \label{subsec:compare_LAB00}
First, problem (\ref{opt:sckt}), for each possible specified measurement $i$, is solved using three methods. The first method is the recursive critical $k$-tuple calculation procedure in \cite{LAB00} implemented by the authors. As in \cite{LAB00}, only critical $k$-tuples containing three basic measurements are sought. The second method is the proposed Min-Cut procedure in Algorithm 1 and Algorithm 2 in Section~\ref{sec:Min_Cut}. The third method is the MILP procedure (with $M = 100$) in Section~\ref{sec:MILP}. The solution of the MILP procedure is used as a reference for accuracy. The IEEE 14-bus benchmark system is analyzed, with two different measurement sets. The first measurement set is from \cite{LAB00} (Section IV, measurement set of Scenario 1), containing measurements from 9 out of 14 buses and 6 out of 19 transmission lines. The second measurement set contains all bus and transmission line measurements, which may be of interest for meter placement. For each $i$, the procedure in \cite{LAB00} (with three basic measurements) and Min-Cut might only provide an overestimate of the cardinality of the sparsest critical $k$-tuple. Table~\ref{tab:LAB_min-cut} lists the percentages of $i$ with overestimation, the average overestimation (over all $i$) and the average relative overestimation (relative to the cardinality of the corresponding sparsest critical $k$-tuple).

Table~\ref{tab:LAB_min-cut} confirms the statement in Section~\ref{subsec:contributions} that the procedure in \cite{LAB00} and Min-Cut should be used in different measurement settings. \cite{LAB00} performs better in a sparse measurement set, while Min-Cut is a better choice in a dense measurement set.
\begin{table}[hbt]
  \begin{center}
    \caption{Comparison between the procedure in \cite{LAB00}, Min-Cut and MILP for the IEEE 14-bus system}
    \begin{tabular}{|c|c|c|c|}
    \hline
    measurement set 1 (sparse metering) & \cite{LAB00} & Min-Cut & MILP \\
    \hline
    solve time (s) & 0.04 & 0.02 & 3.6 \\
    \hline
    percent of meas. with overestimation (\%) & 0 & 93 & 0 \\
    \hline
    average overestimation & 0 & 1.07 & 0 \\
    \hline
    average relative overestimation (\%) & 0 & 75 & 0 \\
    \hline
    \hline
    measurement set 2 (full metering) & \cite{LAB00} & Min-Cut & MILP \\
    \hline
    solve time (s) & 10 & 0.03 & 17 \\
    \hline
    percent of meas. with overestimation (\%) & 26.5 & 0 & 0 \\
    \hline
    average overestimation & 4.1 & 0 & 0 \\
    \hline
    average relative overestimation (\%) & 67.7 & 0 & 0 \\
    \hline
    \end{tabular}
    \label{tab:LAB_min-cut}
  \end{center}
\end{table}

\subsection{The Effect of the Proportion of Line Power Flow Measurements on the Min-Cut Procedure} \label{subsec:lpf_test}
As explained in Section~\ref{sec:Min_Cut}, the Min-Cut procedure for (\ref{opt:sckt}) achieves computation efficiency by approximately counting the injection measurements. Hence, the relative ratio between the line power flow and bus injection measurements affects the approximation quality of the Min-Cut procedure. In this subsection, the relationship between approximation quality and the proportion of transmission line measurements in the network is considered.

The IEEE 14-bus, 57-bus and 118-bus benchmark systems are considered. The network topologies are from MATPOWER \cite{MATPOWER}. For each system, 11 different measurement sets are considered. Each measurement set contains all injection measurements, but the proportions of removed line power flow measurements increase as 0\%, 10\%, \ldots, 100\%. The removed line measurements are randomly chosen. A study similar to the one in Section~\ref{subsec:compare_LAB00} is performed, testing only the proposed Min-Cut and MILP procedures. The above study can be considered as ``one sample'' of a random experiment involving a sequence of 11 measurement sets for each benchmark system. The randomness stems from choice of the removed line flow measurements. To examine the typical phenomena, the above random experiment is repeated five times. Table~\ref{tab:lpf_test_result} shows the mean value (over 5 experiments) of the performance and error statistics similar to those in Table~\ref{tab:LAB_min-cut}.

\begin{table*}[bth]
  \begin{center}
    \caption{Ensemble mean of the solve time and error statistics for the case study in Section~\ref{subsec:lpf_test}}
    \begin{tabular}{|c|c|c|c|c|c|c|c|c|c|c|c|c|}
        \hline
        \multirow{5}{*}{14-bus} & line meas.\ removal (relative to total lines) (\%) & 0 & 10 & 20 & 30 & 40 & 50 & 60 & 70 & 80 & 90 & 100 \\
        \cline{2-13}
        & line meas.\ removal (relative to total meas.) (\%) & 0 & 6 & 12 & 18 & 24 & 29 & 35 & 41 & 47 & 53 & 58 \\
        \cline{2-13}
        & solve time ratio (Min-Cut/MILP) $\times 10^{-4}$ & 90 & 97 & 110 & 140 & 151 & 175 & 262 & 369 & 557 & 924 & 3642 \\
        \cline{2-13}
        & percent of meas.\ with overestimation (\%) & 0 & 0.63 & 1.0 & 1.4 & 3.1 & 5.4 & 6.0 & 5.5 & 6.1 & 6.9 & 86 \\
        \cline{2-13}
        & average overestimation & 0 & 0.1 & 0.2 & 0.3 & 0.5 & 0.7 & 0.96 & 0.97 & 0.85 & 0.8 & 1.0 \\
        \cline{2-13}
        & average relative overestimation (\%) & 0 & 1.7 & 3.1 & 5.7 & 9.0 & 12 & 20 & 21 & 22 & 23 & 50 \\
        \hline
        \hline
        \multirow{5}{*}{57-bus} & line meas.\ removal (relative to total lines) (\%) & 0 & 10 & 20 & 30 & 40 & 50 & 60 & 70 & 80 & 90 & 100 \\
        \cline{2-13}
        & line meas.\ removal (relative to total meas.) (\%) & 0 & 5.8 & 12 & 18 & 23 & 29 & 35 & 40 & 47 & 53 & 58 \\
        \cline{2-13}
        & solve time ratio (Min-Cut/MILP) $\times 10^{-4}$ & 1.5 & 2.1 & 2.7 & 2.4 & 5.9 & 6.6 & 9.5 & 17 & 18 & 36 & 2740 \\
        \cline{2-13}
        & percent of meas.\ with overestimation (\%) & 0 & 1.2 & 1.0 & 1.2 & 0.95 & 2.1 & 3.2 & 3.0 & 2.2 & 0.92 & 96 \\
        \cline{2-13}
        & average overestimation & 0 & 0.4 & 0.6 & 0.65 & 0.8 & 0.6 & 0.8 & 0.8 & 1.0 & 0.2 & 1.1 \\
        \cline{2-13}
        & average relative overestimation (\%) & 0 & 4.7 & 8.9 & 12 & 14 & 11 & 14 & 15 & 23 & 5.0 & 53 \\
        \hline
        \hline
        \multirow{5}{*}{118-bus} & line meas.\ removal (relative to total lines) (\%) & 0 & 10 & 20 & 30 & 40 & 50 & 60 & 70 & 80 & 90 & 100 \\
        \cline{2-13}
        & line meas.\ removal (relative to total meas.) (\%) & 0 & 6.2 & 12 & 18 & 24 & 31 & 37 & 43 & 49 & 55 & 61 \\
        \cline{2-13}
        & solve time ratio (Min-Cut/MILP) $\times 10^{-4}$ & 2.1 & 2.1 & 1.7 & 2.0 & 2.1 & 2.4 & 2.7 & 3.9 & 4.6 & 11 & 1116 \\
        \cline{2-13}
        & percent of meas.\ with overestimation (\%) & 0 & 0.77 & 0.90 & 1.4 & 2.7 & 4.2 & 5.4 & 6.7 & 6.7 & 5.7 & 87 \\
        \cline{2-13}
        & average overestimation & 0 & 1.0 & 1.2 & 1.1 & 1.2 & 1.2 & 1.2 & 1.2 & 1.2 & 1.2 & 1.1 \\
        \cline{2-13}
        & average relative overestimation (\%) & 0 & 13 & 15 & 14 & 17 & 19 & 22 & 24 & 27 & 35 & 54 \\
        \hline
    \end{tabular}
    \label{tab:lpf_test_result}
  \end{center}
\end{table*}

While not seen from Table~\ref{tab:lpf_test_result}, the computation time for Min-Cut remains roughly the same. The increase in solve time ratio (up to about 0.36 when 100\% of line measurements are removed) is due to the decrease in solve time of the MILP procedure. In general, Min-Cut is more efficient than MILP. In terms of approximation error, for up to 90\% of line measurement removal, Min-Cut results in at most 7\% of measurements whose sparsest critical $k$-tuple cardinalities are overestimated (the number is down to 3\% for up to 40\% of line measurement removal). On average, the overestimation is by about 1 measurement with the maximum observed in the experiment being 3 measurements (not shown in Table~\ref{tab:lpf_test_result}). Finally, the average relative overestimation is at worst 35\% (e.g., overestimation by 1 measurement for a critical 3-tuple).

\subsection{The Effect of the Proportion of Injection Measurements and Arbitrary Measurements on the Min-Cut Procedure} \label{subsec:inj_all_test}
In this subsection, the experiment in Section~\ref{subsec:lpf_test} is repeated for the 118-bus benchmark system with a difference in the definition of the measurement sets. Two cases are considered. In the first case, each measurement set contains all line power flow measurements and different proportions of the injection measurements are randomly removed. In the second case, different proportions of arbitrary measurements (injection or line) are randomly removed in a way that the resulted network is still observable. Table~\ref{tab:inj_test_result} lists the relevant statistics for the first case and suggests (from the fourth row) that the Min-Cut procedure is more accurate when the transmission lines are more densely metered. On the other hand, Table~\ref{tab:cmb_test_result} lists the statistics for the second case. In this case, at most $1-118/(118+186) \approx 41\%$ of the measurements can be removed before the network become unobservable. However, random removal of 40\% of the measurements typically results in a unobservable network, and hence the corresponding result is not shown in Table~\ref{tab:cmb_test_result}. Table~\ref{tab:cmb_test_result} confirms again that the Min-Cut procedure is efficient and accurate for relatively dense measurement sets.

\begin{table*}[bth]
  \begin{center}
    \caption{Ensemble mean of the solve time and error statistics for the case study in Section~\ref{subsec:inj_all_test} with varying injection measurement proportion}
    \begin{tabular}{|c|c|c|c|c|c|c|c|c|c|c|c|}
      \hline
      bus meas.\ removal (relative to total buses) (\%) & 0 & 10 & 20 & 30 & 40 & 50 & 60 & 70 & 80 & 90 & 100 \\
      \hline
      bus meas.\ removal (relative to total meas.) (\%) & 0 & 3.9 & 7.8 & 12 & 16 & 19 & 23 & 27 & 31 & 34 & 39 \\
      \hline
      solve time ratio (Min-Cut/MILP) $\times 10^{-4}$ & 2.1 & 3.2 & 6.4 & 9.7 & 18 & 33 & 61 & 108 & 185 & 404 & 1211 \\
      \hline
      percent of meas.\ with overestimation (\%) & 0 & 0.61 & 1.4 & 1.4 & 1.5 & 2.2 & 2.6 & 2.3 & 1.4 & 1.1 & 0 \\
      \hline
      average overestimation & 0 & 1.4 & 1.1 & 1.2 & 1.2 & 1.2 & 1.1 & 1.2 & 1.1 & 0.8 & 0 \\
      \hline
      average relative overestimation (\%) & 0 & 16 & 14 & 15 & 16 & 18 & 19 & 20 & 22 & 15 & 0 \\
      \hline
    \end{tabular}
    \label{tab:inj_test_result}
  \end{center}
\end{table*}

\begin{table*}[bth]
  \begin{center}
    \caption{Ensemble mean of the solve time and error statistics for the case study in Section~\ref{subsec:inj_all_test} with varying arbitrary measurement proportion}
    \begin{tabular}{|c|c|c|c|c|c|c|}
      \hline
      meas.\ removal (relative to total meas.) (\%) & 0 & 10 & 20 & 30 & 40 & 50 \\
      \hline
      solve time ratio (Min-Cut/MILP) $\times 10^{-4}$ & 2.1 & 3.3 & 8.9 & 21 & 110 & 345 \\
      \hline
      percent of meas.\ with overestimation (\%) & 0 & 1.0 & 5.0 & 7.6 & 20 & 35 \\
      \hline
      average overestimation & 0 & 0.9 & 1.0 & 1.1 & 1.1 & 1.4 \\
      \hline
      average relative overestimation (\%) & 0 & 12 & 24 & 28 & 45 & 69 \\
      \hline
    \end{tabular}
    \label{tab:cmb_test_result}
  \end{center}
\end{table*}

\subsection{Time Efficiency of Min-Cut and MILP for Large Networks}
This numerical study investigates the possible advantage of the proposed Min-Cut procedure for the sparsest critical $k$-tuple analysis for larger scale power networks. The networks considered are the IEEE 118-bus, IEEE 300-bus and the Polish 2383-bus systems. The topologies of these networks are obtained using MATPOWER \cite{MATPOWER}. On each transmission line of the benchmark systems, there are two line power flow meters (one from each terminal bus). In addition, all power injections are measured. For each of the benchmark system, problem (\ref{opt:sckt}) is solved using the Min-Cut procedure for all possible specified measurement $i$. For the 118 and 300 bus cases, the Min-Cut procedure is experimentally found to be exact, compared with solving the MILP formulation in (\ref{opt:suoa_MILP}). For the 2383-bus case, only 14 instances of (\ref{opt:suoa_MILP}) are solved. For all these 14 instances Min-Cut also provides the correct estimates. The computation times for using the proposed Min-Cut procedure and solving (\ref{opt:suoa_MILP}) are listed in Table~\ref{tab:sckt_runtime}. This numerical study again assures the efficiency and accuracy of the proposed Min-Cut procedure in the case when all transmission lines and buses are metered.



\subsection{Using Critical $k$-tuple Information for Meter Placement} \label{subsec:meter_placement}
\begin{figure}[!tbh]
  \begin{center}
    \includegraphics[scale=0.4]{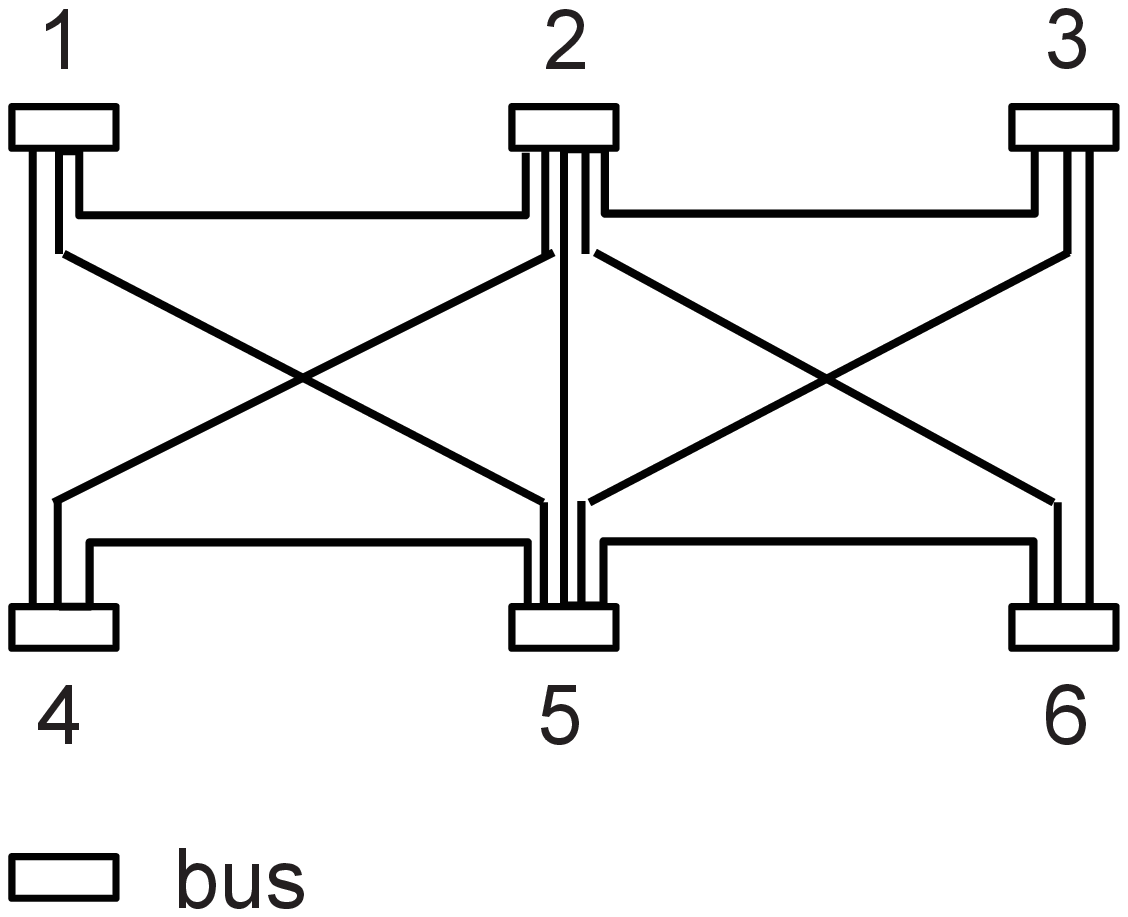}
  \end{center}
  \caption{IEEE 6-bus system.}
  \label{fig:IEEE6bus}
\end{figure}

\begin{figure}[!tbh]
  \begin{center}
    \includegraphics[scale=0.53]{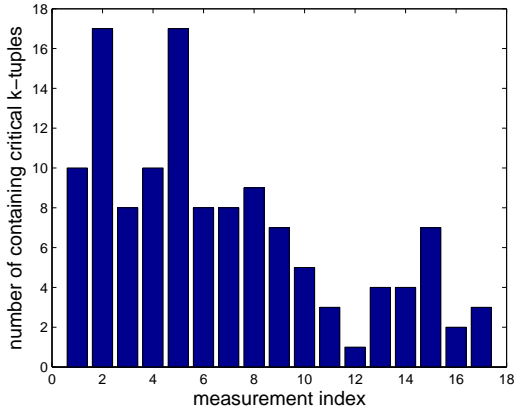}
  \end{center}
  \caption{Number of critical $k$-tuples containing any specific measurement.}
  \label{fig:nckt_j}
\end{figure}

While not the main focus of this paper, a possible use of the critical $k$-tuple information in (\ref{opt:sckt}) is for meter placement. Problem (\ref{opt:sckt}) is solved for all specified measurement $i$ for the IEEE 6-bus benchmark system (see Fig.~\ref{fig:IEEE6bus}). It is assumed that the network is fully metered (with 6 injection and 11 line power flow measurements). The Min-Cut procedure (i.e., Algorithm 1 and Algorithm 2) is used to solve (\ref{opt:sckt}). In total, 17 critical $k$-tuples are found. Then for each measurement $i$, Fig.~\ref{fig:nckt_j} shows the number of critical $k$-tuples containing $i$. However, note that by solving (\ref{opt:sckt}) for all specified measurement $i$, all sparsest critical $k$-tuples are not found. Hence, Fig.~\ref{fig:nckt_j} only shows the lower bounds for the true number of critical $k$-tuples containing $i$. Fig.~\ref{fig:nckt_j} indicates that measurement 12 is probably not important (as far as network observability is concerned), because only one critical $k$-tuple contains it and none of the other 16 critical $k$-tuples will be affected by the removal of measurement 12. This is consistent with the network topology in Fig.~\ref{fig:IEEE6bus}, since measurement 12 is the line power flow measurement between bus 2 and bus 5 (i.e., the two buses with the largest degree). On the other hand, Fig.~\ref{fig:nckt_j} suggests that measurements 2 and 5 (i.e., power injections at bus 2 and bus 5) are definitely important because they are involved in many critical $k$-tuples. This is again consistent with the topology in Fig.~\ref{fig:IEEE6bus} since each of these injection measurements can substitute one of the five line measurements in case any one of them fails. Finally, note that the same analysis here can be carried out for larger scale networks where it becomes less obvious from the topology which measurements are important or unimportant.

\begin{table}[!bth]
  \begin{center}
    \caption{CPU times for solving all instances of (\ref{opt:sckt}) in the 118, 300 and 2383 bus systems}
    \begin{tabular}{|c|c|c|c|}
      \hline
      Method & 118-bus & 300-bus & 2383-bus \\
      \hline
      MILP & 763 sec & 6708 sec & (projected) 5.7 days \\
      \hline
      Min-Cut & 0.30 sec & 1 sec & 31 sec \\
      \hline
    \end{tabular}
    \label{tab:sckt_runtime}
  \end{center}
\end{table}

\section{Conclusion} \label{sec:conclusion}
In conclusion, a version of the sparsest critical $k$-tuple problem is considered. The sparsest critical $k$-tuple is sought for one arbitrarily specified measurement. It is possible to identify the weak points in the power network by listing all measurements which might form critical $k$-tuples with small cardinality, even though this is short of a complete enumeration of all possible sparsest critical $k$-tuples. This paper demonstrates that the studied sparsest critical $k$-tuple problem can be formulated as a MILP problem so that powerful MILP solvers such as CPLEX and Gurobi can be utilized. On the other hand, by using topological network observability results in \cite{KCD80,KJTT10}, a Min-Cut based approximate solution procedure can be derived. The numerical experiment in this paper reveals that the Min-Cut procedure is highly accurate and efficient when there are a significant number of line power flow measurements in the power network. Consequently, Min-Cut should be the first method to attempt (over MILP) in this scenario.

\section*{Appendix}

\subsection*{Proof of Corollary~\ref{rem:equivalence}}
Suppose $\theta^\star$ is an optimal solution to (\ref{opt:suoa}) and $I^\star$ is such that $H(j,:) \theta^\star \neq 0$ if and only if $j \in I^\star$. Then Theorem~\ref{thm:sckt_sa} states that $I^\star$ is a feasible solution to (\ref{opt:sckt}) with the objective value ${\rm card}(I^\star) = {\rm card}(H \theta^\star)$. Now suppose $I^\star$ is not optimal and there exists another feasible solution $\tilde{I}$ of (\ref{opt:sckt}) such that ${\rm card}(\tilde{I}) < {\rm card}(I^\star)$. Then Theorem~\ref{thm:sckt_sa} states that there exists $\tilde{\theta}$, feasible in (\ref{opt:suoa}), such that ${\rm card}(H \tilde{\theta}) \le {\rm card}(\tilde{I}) < {\rm card}(I^\star) = {\rm card}(H \theta^\star)$. This contradicts the assumption that $\theta^\star$ is an optimal solution to (\ref{opt:suoa}). Hence, $\tilde{I}$ does not exist and $I^\star$ is an optimal solution to (\ref{opt:sckt}). To establish the converse, suppose $I^\star$ is an optimal solution to (\ref{opt:sckt}). Then Theorem~\ref{thm:sckt_sa} states that there exists $\theta^\star$ feasible to (\ref{opt:suoa}) such that ${\rm card}(H \theta^\star) \le {\rm card}(I^\star)$. In fact, ${\rm card}(H \theta^\star) = {\rm card}(I^\star)$ and $\theta^\star$ is optimal to (\ref{opt:suoa}). If this is not true, Theorem~\ref{thm:sckt_sa} implies that $I^\star$ would not be optimal.

\bibliographystyle{IEEEtran}
\bibliography{ckt}

\begin{thebibliography}{10}
\providecommand{\url}[1]{#1}
\csname url@samestyle\endcsname
\providecommand{\newblock}{\relax}
\providecommand{\bibinfo}[2]{#2}
\providecommand{\BIBentrySTDinterwordspacing}{\spaceskip=0pt\relax}
\providecommand{\BIBentryALTinterwordstretchfactor}{4}
\providecommand{\BIBentryALTinterwordspacing}{\spaceskip=\fontdimen2\font plus
\BIBentryALTinterwordstretchfactor\fontdimen3\font minus
  \fontdimen4\font\relax}
\providecommand{\BIBforeignlanguage}[2]{{%
\expandafter\ifx\csname l@#1\endcsname\relax
\typeout{** WARNING: IEEEtran.bst: No hyphenation pattern has been}%
\typeout{** loaded for the language `#1'. Using the pattern for}%
\typeout{** the default language instead.}%
\else
\language=\csname l@#1\endcsname
\fi
#2}}
\providecommand{\BIBdecl}{\relax}
\BIBdecl

\bibitem{Monticelli_Wu85}
A.~Monticelli and F.~Wu, ``Network observability: Theory,'' \emph{IEEE
  Transactions on Power Apparatus and Systems}, vol. PAS-104, no.~5, pp.
  1042--1048, May 1985.

\bibitem{KCD80}
G.~Krumpholz, K.~Clements, and P.~Davis, ``Power system observability: A
  pratical algorithm using network topology,'' \emph{IEEE Transactions on Power
  Apparatus and Systems}, vol. PAS-99, no.~4, pp. 1534--1542, July/Aug 1980.

\bibitem{Abur_Exposito_SEbook}
A.~Abur and A.~Exp\'{o}sito, \emph{Power System State Estimation}.\hskip 1em
  plus 0.5em minus 0.4em\relax Marcel Dekker, Inc., 2004.

\bibitem{Monticelli_SEbook}
A.~Monticelli, \emph{State Estimation in Electric Power Systems A Generalized
  Approach}.\hskip 1em plus 0.5em minus 0.4em\relax Kluwer Academic Publishers,
  1999.

\bibitem{AAG08}
M.~de~Almeida, E.~Asada, and A.~Garcia, ``On the use of gram matrix in
  observability analysis,'' \emph{IEEE Transactions on Power Systems}, vol.~23,
  no.~1, pp. 249 --251, feb. 2008.

\bibitem{Gou06}
B.~Gou, ``Jacobian matrix-based observability analysis for state estimation,''
  \emph{IEEE Transactions on Power Systems}, vol.~21, no.~1, pp. 348 -- 356,
  feb. 2006.

\bibitem{Pruneda2010277}
R.~Pruneda, C.~Solares, A.~Conejo, and E.~Castillo, ``An efficient algebraic
  approach to observability analysis in state estimation,'' \emph{Electric
  Power Systems Research}, vol.~80, no.~3, pp. 277 -- 286, 2010.

\bibitem{solares:336}
C.~Solares, A.~Conejo, E.~Castillo, and R.~Pruneda, ``Binary-arithmetic
  approach to observability checking in state estimation,'' \emph{IET
  Generation, Transmission \& Distribution}, vol.~3, no.~4, pp. 336--345, 2009.

\bibitem{CCPS06}
E.~Castillo, A.~Conejo, R.~Pruneda, and C.~Solares, ``Observability analysis in
  state estimation: a unified numerical approach,'' \emph{IEEE Transactions on
  Power Systems}, vol.~21, no.~2, pp. 877 -- 886, may 2006.

\bibitem{Monticelli_Wu_Yen86}
A.~Monticelli, F.~Wu, and M.~Yen, ``Multiple bad data identification for state
  estimation by combinatorial optimization,'' \emph{IEEE Transactions on Power
  Delivery}, vol. PWRD-1, no.~3, pp. 361--369, July 1986.

\bibitem{Clements_Davis86}
K.~Clements and P.~Davis, ``Multiple bad data detectability and identification:
  A geometric approach,'' \emph{IEEE Transactions on Power Delivery}, vol.
  PWRD-1, no.~3, pp. 355--360, July 1986.

\bibitem{MVCRP85}
T.~Van~Cutsem, M.~Ribbens-Pavella, and L.~Mili, ``Bad data identification
  methods in power system state estimation-a comparative study,'' \emph{IEEE
  Transactions on Power Apparatus and Systems}, vol. PAS-104, no.~11, pp. 3037
  --3049, 1985.

\bibitem{KC91}
G.~Korres and G.~Contaxis, ``Identification and updating of minimally dependent
  sets of measurements in state estimation,'' \emph{IEEE Transactions on Power
  Systems}, vol.~6, no.~3, pp. 999 --1005, aug 1991.

\bibitem{AAG09}
M.~de~Almeida, E.~Asada, and A.~Garcia, ``Identifying critical sets in state
  estimation using gram matrix,'' in \emph{PowerTech, 2009 IEEE Bucharest}, 28
  2009-july 2 2009, pp. 1 --5.

\bibitem{AH86}
M.~Ayres and P.~H. Haley, ``Bad data groups in power system state estimation,''
  \emph{IEEE Transactions on Power Systems}, vol.~1, no.~3, pp. 1 --7, aug.
  1986.

\bibitem{CKD81}
K.~Clements, G.~Krumpholz, and P.~Davis, ``Power system state estimation
  residual analysis: An algorithm using network topology,'' \emph{Power
  Apparatus and Systems, IEEE Transactions on}, vol. PAS-100, no.~4, pp. 1779
  --1787, april 1981.

\bibitem{LondonJr2004583}
{\rm J.B.A. London Jr and A.S. Bretas and N.G. Bretas}, ``Algorithms to solve
  qualitative problems in power system state estimation,'' \emph{International
  Journal of Electrical Power \& Energy Systems}, vol.~26, no.~8, pp. 583 --
  592, 2004.

\bibitem{LAB00}
J.~London, J.B.A., L.~Alberto, and N.~Bretas, ``Network observability:
  identification of the measurements redundancy level,'' in \emph{Power System
  Technology, 2000. Proceedings. PowerCon 2000. International Conference on},
  vol.~2, 2000, pp. 577 --582 vol.2.

\bibitem{CCPSM08}
E.~Castillo, A.~Conejo, R.~Pruneda, C.~Solares, and J.~Menendez, ``m-k robust
  observability in state estimation,'' \emph{IEEE Transactions on Power
  Systems}, vol.~23, no.~2, pp. 296 --305, may 2008.

\bibitem{GA01}
B.~Gou and A.~Abur, ``An improved measurement placement algorithm for network
  observability,'' \emph{IEEE Power Engineering Review}, vol.~21, no.~10,
  p.~61, oct. 2001.

\bibitem{KJTT10}
O.~Kosut, L.~Jia, R.~Thomas, and L.~Tong, ``Malicious data attacks on smart
  grid state estimation: Attack strategies and countermeasures,'' in \emph{IEEE
  SmartGridComm}, 2010.

\bibitem{LRN09}
Y.~Liu, M.~Reiter, and P.~Ning, ``False data injection attacks against state
  estimation in electric power grids,'' in \emph{16th ACM Conference on
  Computer and Communication Security}, New York, NY, USA, 2009, pp. 21--32.

\bibitem{BRWKNO10}
R.~Bobba, K.~Rogers, Q.~Wang, H.~Khurana, K.~Nahrstedt, and T.~Overbye,
  ``Detecting false data injection attacks on dc state estimation,'' in
  \emph{the First Workshop on Secure Control Systems, CPSWEEK 2010}, 2010.

\bibitem{IR-EE-RT_2010:071}
A.~Teixeira, S.~Amin, H.~Sandberg, K.~H. Johansson, and S.~Sastry, ``Cyber
  security analysis of state estimators in electric power systems,'' in
  \emph{Proceedings {IEEE} Conference on Decision and Control}, dec 2010.

\bibitem{DS_SGC2010}
G.~Dan and H.~Sandberg, ``Stealth attacks and protection schemes for state
  estimators in power systems,'' in \emph{IEEE SmartGridComm}, 2010.

\bibitem{SSJ_CDC2011_mincut}
{\rm K.C.~Sou and H.~Sandberg and K.H.~Johansson}, ``Electric power network
  security analysis via minimum cut relaxation,'' in \emph{IEEE Conference on
  Decision and Control}, December 2011, report version:
  \url{https://eeweb01.ee.kth.se/upload/publications/reports/2011/IR-EE-RT\_2011\_089.pdf}.

\bibitem{BT97}
J.~Tsitsiklis and D.~Bertsimas, \emph{Introduction to Linear
  Optimization}.\hskip 1em plus 0.5em minus 0.4em\relax Athena Scientific,
  1997.

\bibitem{Stoer:1997:SMA:263867.263872}
M.~Stoer and F.~Wagner, ``A simple min-cut algorithm,'' \emph{J. ACM}, vol.~44,
  pp. 585--591, July 1997.

\bibitem{FF_MAX-FLOW}
L.~Ford and D.~Fulkerson, ``Maximal flow through a network,'' \emph{Canadian
  Journal of Mathematics}, vol.~8, pp. 399--404, 1956.

\bibitem{springerlink:10.1007/BFb0120902}
J.-C. Picard and M.~Queyranne, ``On the structure of all minimum cuts in a
  network and applications,'' in \emph{Combinatorial Optimization II}, ser.
  Mathematical Programming Studies, 1980, vol.~13, pp. 8--16.

\bibitem{Schrage_Baker_1978}
L.~Schrage and K.~R. Baker, ``Dynamic programming solution of sequencing
  problems with precedence constraints,'' \emph{Operations Research}, vol.~26,
  no.~3, pp. pp. 444--449, 1978.

\bibitem{Lawer_1979}
\BIBentryALTinterwordspacing
E.~L. Lawler, ``\BIBforeignlanguage{en}{Efficient {Implementation} {Of}
  {Dynamic} {Programming} {Algorithms} {For} {Sequencing} {Problems} :
  (preprint)},'' CWI Technical Report Stichting Mathematisch Centrum.
  Mathematische Besliskunde-BW 106/79, 1979. [Online]. Available:
  \url{http://oai.cwi.nl/oai/asset/9663/9663A.pdf}
\BIBentrySTDinterwordspacing

\bibitem{Cunningham85submodularfunction}
W.~H. Cunningham, ``On submodular function minimization,''
  \emph{Combinatorica}, vol.~5, pp. 185--192, 1985.

\bibitem{Mccormick07submodularfunction}
S.~T. Mccormick, ``Submodular function minimization based on chapter 7 of the
  handbook on discrete optimization [54] version 3,'' 2007.

\bibitem{BGT81}
R.~G. Bland, D.~Goldfarb, and M.~J. Todd, ``The ellipsoid method: A survey,''
  in \emph{Operations Research}, 1981.

\bibitem{Iwata:2009:SCA:1496770.1496903}
S.~Iwata and J.~B. Orlin, ``A simple combinatorial algorithm for submodular
  function minimization,'' in \emph{Proceedings of the twentieth Annual
  ACM-SIAM Symposium on Discrete Algorithms}, ser. SODA '09, 2009, pp.
  1230--1237.

\bibitem{Mallat93matchingpursuit}
S.~Mallat and Z.~Zhang, ``Matching pursuit with time-frequency dictionaries,''
  \emph{IEEE Transactions on Signal Processing}, vol.~41, pp. 3397--3415, 1993.

\bibitem{STJ_SCS2010}
{\rm H. Sandberg, A. Teixeira and K.H. Johansson}, ``On security indices for
  state estimators in power networks,'' in \emph{First Workshop on Secure
  Control Systems, CPSWEEK 2010}, 2010.

\bibitem{LASSO}
R.~Tibshirani, ``{\rm Regression shrinkage and selection via the LASSO},''
  \emph{J. Royal. Statist. Soc B}, vol.~58, no.~1, pp. 267--288, 1996.

\bibitem{CPLEX}
``Cplex,''
  \url{http://www-01.ibm.com/software/integration/optimization/cplex-optimizer/}.

\bibitem{Gurobi}
``Gurobi,'' \url{http://www.gurobi.com/}.

\bibitem{Gleich06contentsmatlab}
D.~Gleich, ``{\rm Contents Matlab BGL v4.0},''
  \url{http://www.stanford.edu/\textasciitilde dgleich/programs/matlab\_bgl/},
  2006.

\bibitem{2002:BGL:504206}
\emph{The boost graph library: user guide and reference manual}.\hskip 1em plus
  0.5em minus 0.4em\relax Boston, MA, USA: Addison-Wesley Longman Publishing
  Co., Inc., 2002.

\bibitem{Gurobi_Mex}
W.~Yin, ``{\rm Gurobi Mex: A MATLAB interface for Gurobi},''
  \url{http://convexoptimization.com/wikimization/index.php/gurobi\_mex}.

\bibitem{MATPOWER}
R.~Zimmerman, C.~Murillo-S\'{a}nchez, and R.~Thomas, ``{\rm MATPOWER
  Steady-State Operations, Planning and Analysis Tools for Power Systems
  Research and Education},'' \emph{IEEE Transacations on Power Systems},
  vol.~26, no.~1, pp. 12--19, 2011.

\end{thebibliography}

\end{document}